\documentclass[twocolumn,english]{revtex4}
\usepackage[T1]{fontenc}
\usepackage[latin9]{inputenc}
\usepackage{color}
\usepackage{amsmath}
\usepackage{graphicx}
\usepackage{amssymb}

\makeatletter
\@ifundefined{textcolor}{}
{%
 \definecolor{BLACK}{gray}{0}
 \definecolor{WHITE}{gray}{1}
 \definecolor{RED}{rgb}{1,0,0}
 \definecolor{GREEN}{rgb}{0,1,0}
 \definecolor{BLUE}{rgb}{0,0,1}
 \definecolor{CYAN}{cmyk}{1,0,0,0}
 \definecolor{MAGENTA}{cmyk}{0,1,0,0}
 \definecolor{YELLOW}{cmyk}{0,0,1,0}
 }

\usepackage{bbm}
\def\id{\mathbbm{1}}
\def\vec#1{\mathbf #1}

\makeatother

\usepackage{babel}

\begin{document}
\global\long\def\bra#1{\langle#1|}

\global\long\def\ket#1{|#1\rangle}

\global\long\def\ave#1{\langle#1\rangle}

\global\long\def\id{\mathbb{I}}

\title{Decoherence of a quantum gyroscope}

\author{Olivier Landon-Cardinal}
\affiliation{D\'epartement de Physique, Universit\'e de Sherbrooke,
Sherbrooke, Qu\'ebec, J1K 2R1 Canada}

\author{Richard MacKenzie}
\affiliation{Groupe de physique des particules, Universit\'e de Montr\'eal,
C.P. 6128, Succ. Centre-ville, Montr\'eal, Qu\'ebec, H3C 3J7 Canada}

\date{\today}

\begin{abstract}
We study the behavior of a quantum gyroscope, that is, a quantum system which
singles out a direction in space in order to measure certain properties of
incoming particles such as the orientation of their spins. 
We show that repeated Heisenberg interactions of the gyroscope with
several incoming spin-1/2 particles provides a simple model of decoherence
which exhibits both relaxation and dephasing. Focusing
on the semiclassical limit, we derive equations of motion for the evolution of a
coherent state and investigate the evolution of a superposition
of such states. While a coherent state evolves on a timescale given by the
classical ratio of the angular momentum of the gyroscope to that of the incoming
particles, dephasing acts on a much shorter timescale that depends only on the
angular difference of the states in the superposition.\end{abstract}
\maketitle

\section{Introduction}


A quantum gyroscope is a quantum system that singles out a direction in space.
It allows for the measurement of the spin of incoming particles through the
measurement of the total spin of both the gyroscope and the incoming particle.
This is one important example of the more general theory of quantum reference
frames~\cite{BRS07}. While the gyroscope allows the measurement of spin along
its direction, it will necessarily degrade with time. We will
briefly review work that has been done in that direction before
revisiting the same model but not as a quantum reference frame, but as a simple
model of decoherence.

In~\cite{BRS+06}, Bartlett, {\it et
al.} studied a model in which the reference
frame is represented by a spin $\ell$ in a polarized state. A succession of
unpolarized spin-1/2 particles (all initially in the same state) interact with
the reference, the interaction chosen being a measurement of the total spin,
after which the particle is discarded and the result of the measurement is
ignored. The number of interactions acts as a time parameter, and they showed
that this ``quantum gyroscope'' would degrade on a time scale proportional to
$\ell^2$. A similar model was studied by Poulin and Yard~\cite{PY07}, who
considered the spin-1/2 particles to be partially polarized along a fixed axis.
They found that, in addition to degradation, the reference spin drifted towards
the direction of the polarization of the spin-1/2 particles on a time scale {\em
linear} in $\ell$, making it in a sense a more dangerous effect than degradation
since, for large $\ell$, it occurs on a faster time scale. They concluded that
the gyroscope would evolve semiclassically and be useful to measure
spin-$\frac{1}{2}$ particles along the drifted direction rather than its
original polarization. In~\cite{AJR10}, Ahmadi, {\it et al.} analyzed the
evolution of the gyroscope after repeated measurements and showed that by
retaining the outcomes of the measurements, one could correct the drift of the
quantum gyroscope.

In this paper, building on previous work \cite{LM}, we use a similar
model of quantum gyroscope but focus on its decoherence 
in order to analyze the transition between quantum
and semiclassical behavior. From this new point of view, the model
is promising since it exhibits a sharp transition between these two
regimes. The setup we study is representative of many decoherence
processes: a quantum system placed in a semiclassical state (the gyroscope)
interacts with a large number of incoming particles which together
form its environment. However, unlike most decoherence models, the transfer
of information to the environment occurs since each incoming particle
interacts successively with the quantum gyroscope. 

Another important motivation
for this work is linked to a strategy recently devised to reduce decoherence in
quantum dot spin qubits. In these systems, it was realized that the main
source of decoherence is the hyperfine interaction between the spin qubit and
the Overhauser field $\mathbf{h}$ resulting from all the nuclear spins in the
substrate. The corresponding Hamiltonian is $H=\mathbf{S}\cdot\mathbf{h}$ where
$\mathbf{S}$ is the spin operator of the spin qubit.
In the presence
of an external magnetic field, the essential contribution to the interaction
Hamiltonian is due to the component of the spin along the direction of the
magnetic field. The interaction thus reduces to a tensor product. For such a
Hamiltonian preparing an eigenstate of the Overhauser field $\mathbf{h}$ will
greatly reduce decoherence~\cite{IKT+03,CL04,CPZ05}, and will even suppress it if the
prepared state is also an eigenstate of the self-evolution of the nuclear
spins~\cite{LM09}.
Therefore, recent experiments have tried to polarize the nuclear
bath~\cite{PTJ+08,RTP+08} or lock its polarization~\cite{VNK+09,XYS+09}.
Preparing the initial state of the nuclei in a fully polarized state would
be exactly the situation we consider in this paper if the nuclear spins
were restricted to be in a symmetric state and could therefore be considered
as one particle with large spin $\ell$. Therefore, the model we use can be
considered a toy-model that describes the relaxation and dephasing of the
nuclear spins when they interact with stray electrons.

In the next section, we describe the model studied which analyzes the effect on
the quantum gyroscope of a succession of interactions with spin-1/2 particles.
This effect can be written as a quantum channel which, with a particular choice
of the interaction time, is very similar to the quantum channel where a
joint measurement of the total angular momentum is performed~\cite{PY07}.
In Section III we obtain equations for the orientation and magnitude of the
average value of the angular momentum of the gyroscope in the semiclassical
limit, and discuss quantum effects. In Section IV we discuss the evolution of a
coherent state and a superposition of two such states. In the fifth section we
argue that coherent states should minimize purity loss but nonetheless suffer a
significant purity loss initially. In the final section, we discuss what can be
learned about the state of the reference by measuring the spin of the
interacting particles rather than simply discarding them.


\section{The model}

Our model is summed up in the following timeline. At $t=0$, a spin-$\frac{1}{2}$
particle $\mathcal{S}$ in a state described by a density matrix $\xi$
starts to interact with the spin-$\ell$ gyroscope $\mathcal{R}$
according to a Heisenberg Hamiltonian $\mathcal{H}_{\mathcal{RS}}=\mathbf{L\cdot
S}$
where $\mathbf{L}$ (resp.~$\mathbf{S}$) is the angular momentum
operator of the gyroscope (resp.~of the particle). At $t=\tau$,
the interaction ceases and the particle is discarded. This defines
a quantum channel $\mathcal{E}$ mapping the state of the gyroscope
$\mathcal{R}$ before interacting with the particle, represented by a density
matrix $\rho$ of dimension
$d=2\ell+1$, to its state after the interaction. Immediately thereafter, a
second particle, also in state $\xi$, starts to interact
with the gyroscope, the Hamiltonian and duration as above. This process is then
repeated $n$ times. 

This model has also been discussed by Ahmadi, Jennings and Rudolph
recently~\cite{AJR10}; their main goal was to see if the drift of the gyroscope
frame
noted in~\cite{PY07} could be eliminated. They considered and compared several
possible strategies to correct for this drift.

We can write a discrete iterative equation for the state of the gyroscope. If
after $n$ interactions it is in the state $\mathcal{E}^{n}(\rho)$ (where
$\mathcal{E}^{0}(\rho)=\rho$), then \begin{eqnarray}
\mathcal{E}^{n+1}(\rho) & \equiv &
\rho((n+1)\tau)\label{eq:channel_definition}\\
 & = &
\mbox{Tr}_{\mathcal{S}}\left[e^{-i\mathcal{H}_{\mathcal{RS}}\tau}\left(\mathcal{
E}^{n}(\rho)\otimes\xi\right)e^{+i\mathcal{H}_{\mathcal{RS}}\tau}\right]
\label{eq:iteration_expression}\end{eqnarray}

The evolution operator $e^{-i\mathcal{H}_{\mathcal{RS}}\tau}$ can
be expressed analytically by noting that $\forall\ k,\ (\mathbf{L\cdot
S})^{k}=a_{k}\id+b_{k}(\mathbf{L\cdot S)}$.
The coefficients $a_{k}$ and $b_{k}$ obey a set of coupled linear recurrence
relations 
\begin{eqnarray}
a_{k+1} & = & \frac{\ell(\ell+1)}{4}b_{k}\\
b_{k+1} & = & a_{k}-\frac{b_{k}}{2}\mbox{.}
\end{eqnarray}
Thus, $e^{-i\mathcal{H}_{\mathcal{RS}}\tau}=a(\tau)\id+b(\tau)\mathbf{L\cdot
S}$, where the functions $a(\tau),\ b(\tau)$ are given by
\begin{eqnarray}
a(\tau)&=&\frac{d+1}{2d}e^{-i\frac{d-1}{4}\tau}+\frac{d-1}{2d}e^{i\frac{d+1}{4}
\tau},\\
b(\tau)&=&\frac{2}{d}\left(e^{-i\frac{d-1}{4}\tau}-e^{i\frac{d+1}{4}\tau}
\right).
\end{eqnarray}
This technique, using a decomposition of the $k^{th}$ power of the
interaction Hamiltonian in a finite sum of operators, might be used
for interaction with spins of higher dimension.

Finally, the calculation of \eqref{eq:iteration_expression} for one
particle interacting with the gyroscope ($n=0$) yields~\cite{LM,AJR10}
\begin{eqnarray}
\mathcal{E}(\rho) & = & \left(\cos^{2}\frac{\tau d}{4}+\frac{\sin^{2}\frac{\tau
d}{4}}{d^{2}}\right)\rho+\frac{4}{d^{2}}\sin^{2}\frac{\tau
d}{4}\{\rho,\mathbf{L}\cdot\langle\mathbf{S}\rangle\}\nonumber \\
 &  & +\frac{16}{d^{2}}\sin^{2}\frac{\tau
d}{4}\mbox{Tr}_{\mathcal{S}}\mathbf{L\cdot
S}\left(\rho\otimes\xi\right)\mathbf{L\cdot S}\nonumber \\
 &  & +\frac{2}{d}i\sin\frac{\tau
d}{2}[\rho,\mathbf{L}\cdot\langle\mathbf{S}\rangle]\label{eq:channel}
\end{eqnarray}
This quantum channel, for interaction time $\tau=\frac{\pi}{d}$,
yields a simliar quantum channel than the one based on measuring the
total angular momentum described in~\cite{PY07}, except for
the last term (absent in~\cite{PY07}),
which corresponds to a precession around the axis of preferred polarization
$\langle\mathbf{S}\rangle$. Thus, up to a slight rotation,
interaction and joint measurement imply the
same evolution of the gyroscope. It thus seems that this interaction time
maximize the extraction of information on the incoming particles and we thus
expect it to maximally degrade the gyroscope.

\begin{figure}
\begin{centering}
\includegraphics[width=0.8\columnwidth]{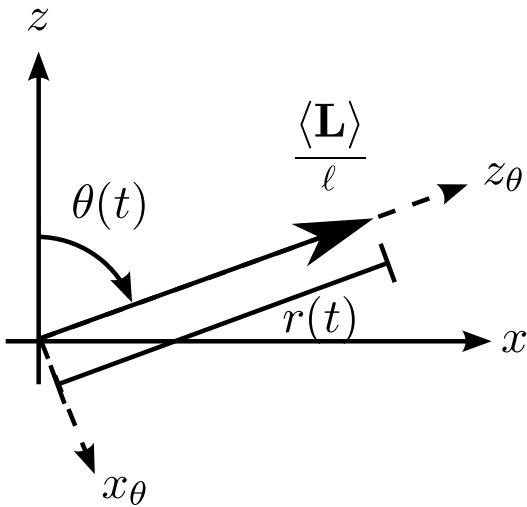}
\par\end{centering}
\caption{The axes are defined such that the gyroscope points in the $\theta(t)$
direction with respect to the $z$ axis and the polarization of the gyroscope is
$\ell r(t)$. $\theta(t)$ naturally defines a rotated system of axes, labeled
$x_\theta$ and $z_\theta$.}
\label{fig:axis-definition}
\end{figure}

In order to pursue the analysis further, we will choose
coordinates as in~\cite{PY07} such that the $z$
axis corresponds to the polarization of the incoming particles and the $x$ axis
is chosen so that initially $\langle\mathbf{L}\rangle$ lies in the $xz$
plane. We assume $\xi=\frac{\id}{2}+2\langle S_{z}\rangle S_{z}$. Rather than
working with operators $\mathbf{L}$ it is convenient
to rotate the axes by an angle $\theta(t)$ around the $y$ axis such
that $L_{x}^{\theta}=\cos\theta\, L_{x}-\sin\theta\, L_{z}$ and
$L_{z}^{\theta}=\sin\theta\, L_{x}+\cos\theta\, L_{z}$.
The angle $\theta(t)$, represented in Fig. \ref{fig:axis-definition}, is chosen
so that $\langle L_{x}^{\theta(t)}(t)\rangle=0$
and $\langle L_{z}^{\theta(t)}(t)\rangle=\ell r(t)$, where $0 \leq r(t)\leq 1$.

%

\section{Semiclassical equations of motion}

\subsection{Systematic method}

Equations of evolution for $\theta(t)$ and $r(t)$ would allow us to determine
the quasiclassical evolution of the gyroscope. An equation for
$\theta(t)$ was already derived in \cite{PY07}, but not
a corresponding equation for $r(t)$.
Here, we provide a systematic method that allows us to derive an equation
for $r(t)$ which (like that for $\theta(t)$) is valid to order $O(1/\ell^{2})$.
More specifically,
we show that macroscopic expectation values $\langle L_{x}^{\theta}\rangle$
and $\langle L_{z}^{\theta}\rangle$ at time $t+1$ can be expressed
analytically in terms of $\langle\left(L_{x}^{\theta}\right)^{2}\rangle$,
$\langle\left(L_{y}^{\theta}\right)^{2}\rangle$,
$\langle\left(L_{z}^{\theta}\right)^{2}\rangle$
and $\langle L_{x}^{\theta} L_{y}^{\theta} \rangle$
at time $t$, where $\theta$ is the angle corresponding to the choice
of axes at time $t$. To do so, we write the quantum channel \eqref{eq:channel}
as a function of the $L_i^\theta$ by using the Kraus form provided in
\cite{PY07}. 
\begin{eqnarray}
\mathcal{E}(\rho) & = &
\frac{d^{2}+1}{2d^{2}}+\frac{2}{d^{2}}\sum_{i=1}^{3}L_{i}^{\theta}\rho
L_{i}^{\theta}\nonumber \\
 &  & +\frac{4}{d^{2}}i\ave{S_{z}}\sin\theta\left(L_{y}^{\theta}\rho
L_{z}^{\theta}-L_{z}^{\theta}\rho L_{y}^{\theta}\right)\nonumber \\
 &  & +\frac{4}{d^{2}}i\ave{S_{z}}\cos\theta\left(L_{y}^{\theta}\rho
L_{x}^{\theta}-L_{x}^{\theta}\rho L_{y}^{\theta}\right)\nonumber \\
 &  & +\frac{2}{d^{2}}\ave{S_{z}}\cos\theta\left(L_{z}^{\theta}\rho+\rho
L_{z}^{\theta}\right)\nonumber \\
 &  & -\frac{2}{d^{2}}\ave{S_{z}}\sin\theta\left(L_{x}^{\theta}\rho+\rho
L_{x}^{\theta}\right)\end{eqnarray}
Calculation of
$\ave{L_{i}^{\theta}(t+1)}=\mbox{Tr}\left[L_{i}^{\theta}(t)\mathcal{E}
(\rho)\right]$
yields

\begin{eqnarray}
\ave{L_{x}^{\theta}(t+1)} & = &
-\frac{4}{d^{2}}\ave{S_{z}}\cos\theta\ave{L_{x}^{\theta}L_{z}^{\theta}}
\nonumber \\
 &  &
-\frac{4}{d^{2}}\ave{S_{z}}\sin\theta\left((\ell(\ell+1)-\ave{\left(L_{x}^{
\theta}\right)^{2}}\right) \\
\ave{L_{z}^{\theta}(t+1)} & = &
(1-\frac{2}{d^{2}})\ave{L_{z}^{\theta}}+\frac{4}{d^{2}}\ave{S_{z}}\sin\theta\ave
{L_{x}^{\theta}L_{z}^{\theta}}\nonumber \\
 &  &
+\frac{4}{d^{2}}\ave{S_{z}}\cos\theta\left(\ell(\ell+1)-\ave{\left(L_{z}^{
\theta}\right)^{2}}\right)\end{eqnarray}

These expressions will allow us to derive equations of motion in
the semiclassical regime.


\subsection{Evolution from a semiclassical state}

Now, suppose that $\rho$ is in a semiclassical state, i.e., a state
that is close enough to an eigenstate of $L_{z}^{\theta}$. In this regime, we
assume 
\begin{equation}
L_{z}^{\theta}\rho\simeq\ell r\rho 
\label{eq:eigenstate-assumption} 
\end{equation}
which implies that $\langle\left(L_{x}^{\theta}\right)^{2}\rangle\simeq \ell^2
r^2$ and $\langle L_{x}^{\theta} L_{y}^{\theta} \rangle \simeq 0$.
Furthermore, we are interested in the semiclassical limit $\ell\gg1$.
In that limit, numerical simulations show that
$\langle\left(L_{x}^{\theta}\right)^{2}\rangle\simeq\langle\left(L_{y}^{\theta}
\right)^{2}\rangle\simeq\left(\ell(\ell+1)-\ell^{2}r^{2}\right)/2$. The average
values thus
reduce to \begin{eqnarray}
\ave{L_{x}^{\theta}(t+1)} & \simeq & -\frac{1+r^{2}}{2} \ave{S_{z}}\sin\theta 
\\
\ave{L_{z}^{\theta}(t+1)} & \simeq & \ell r+(1-r^{2})\ave{S_{z}}\cos\theta
\end{eqnarray}

We can then derive the equations of motion on the macroscopic variables
$r(t),\,\theta(t)$ by using 
\begin{eqnarray}
\dot{\theta} & \simeq & \ave{L_{x}^{\theta}(t+1)}/\ave{L_{z}^{\theta}(t+1)}
\label{eq:thetaexpvalues}\\
2r\dot{r} & = & r^{2}(t+1)-r^{2}(t) \label{eq:rexpvalues}
\end{eqnarray}

The final result is 
\begin{eqnarray}
\dot{r} & = & \lambda\left(1-r^{2}(t)\right)\cos\theta(t)+O(1/\ell^{2})
\label{eq:semiclassical1} \\
\dot{\theta} & = & -\lambda \frac{1+r^{2}(t)}{2r(t)}\sin\theta(t)+O(1/\ell^{2})
\label{eq:semiclassical2}
\end{eqnarray} 
where $\lambda={\langle S_{z}\rangle}/{\ell}$ sets the \emph{time
scale of the evolution}. 

Equations \eqref{eq:semiclassical1} and \eqref{eq:semiclassical2} provide a
clear picture of the semiclassical evolution of the gyroscope. This set of
equations improves on the work in~\cite{PY07} which only derived heuristically a
differential equation for $\theta(t)$ valid in the limit $r=1$. Our method is
more systematic and allows us
to derive an equation for $r(t)$. Furthermore, in~\cite{PY07}, it
was argued that $r$ remains close to $1$ based on numerical observations. 
We strengthen this statement since Eq. \eqref{eq:semiclassical2}
provides an analytic confirmation
that $r$ does indeed remain close to $1$ when $r(0)\simeq1$ and
$\theta\neq\pi$.

The striking feature of the set of coupled equations \eqref{eq:semiclassical1}
and
\eqref{eq:semiclassical2} 
is that the time scale is entirely set by the ratio $\lambda$ of
the angular momentum of the incoming particle $\ave{S_{z}}$ and the
angular momentum of the gyroscope $\ell$. Thus, the semiclassical
evolution of the quantum gyroscope only depends on three parameters: $\ell$
and $r$ that characterize the semiclassical state of the gyroscope
and the ratio $\lambda$. This set of parameters is extremely small
when compared to the coefficients necessary to express the states
of the gyroscope and the incoming particle as vectors in their respective
Hilbert spaces. From an operational point of view the semiclassical
evolution is entirely described by these three parameters and does not
depend on the microscopic details of the gyroscope and the incoming
particle.

\subsection{Quantum effects for a finite gyroscope}

\begin{figure}
\begin{centering}
\includegraphics[width=1\columnwidth]{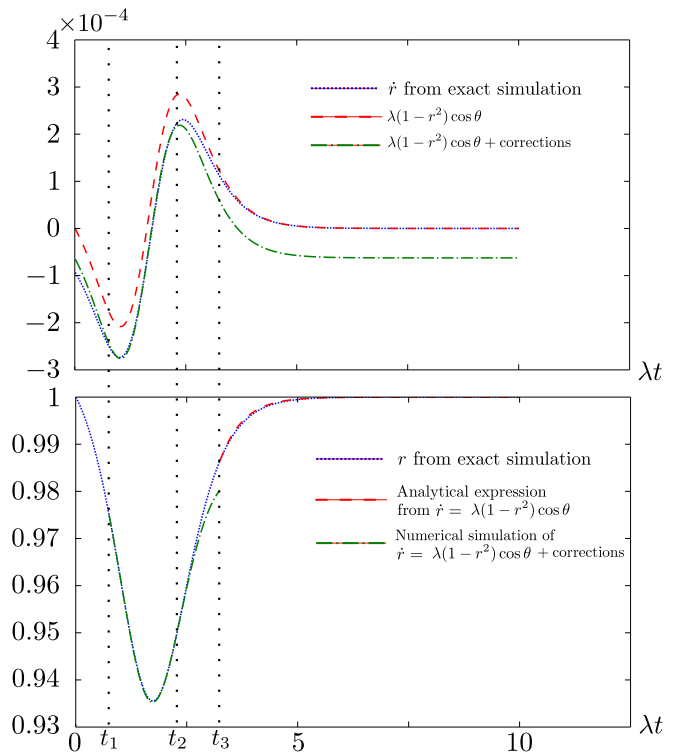}
\end{centering}
\caption{Comparison between the evolution of the polarization $r(t)$ by brute
force numerical simulation of the quantum channel and the analytical solution.
Four regions appear. The initial drop ($t<t_1$) and the crossover region
($t_2<t<t_3$) are not described by the semiclassical eq. of motions and
seems to require a full quantum treatment. On the contrary, the polarization
decrease ($t_1<t<t_2$) and increase ($t_3<t$) are described very accurately by 
semiclassical equations.}
\label{fig:evo-analytical}
\end{figure}

Numerically, one notices that a gyroscope initially fully polarized will
undergo an initial polarization drop as can be seen on Fig.
\ref{fig:evo-analytical}. This phenomenon might shed new light on recent
experiments involving quantum dots~\cite{PTJ+08,RTP+08} where the nuclear bath
reached a level of polarization lower than had been expected.  
The semiclassical equation of motion \eqref{eq:semiclassical1} alone fails to
explain why the polarization undergoes an initial drop. Indeed, for a quantum
gyroscope initially fully polarized ($r=1$), eq. \eqref{eq:semiclassical1}
breaks down because the evolution of the polarization is dominated by
corrections to the otherwise-dominant term. A more careful derivation yields  
\begin{equation}
\dot{r}  \simeq \lambda\left(1-r^{2}\right)\cos\theta -
\frac{1+\ave{S_z}^2}{2\ell^2} r - \frac{\lambda^2}{2} \sin^2\theta r^3 
\label{eq:semiclassical-with-corrections} 
\end{equation}
in the limit where $\ell \gg 1$ and for $\theta > \pi /2$. When the gyroscope is
fully polarized and $\theta$ is close to $\pi$, the second term of
\eqref{eq:semiclassical-with-corrections} dominates and the polarization decays
exponentially
 which explains the initial drop that brings the polarization to the form
$1-\varepsilon$. The first term of \eqref{eq:semiclassical-with-corrections}
takes over but vanishes again for $\theta=\pi/2$. Thus corrections are
responsible for both the initial drop and the precise position of the minimum of
the polarization.

For small $\theta$ eq \eqref{eq:semiclassical1} is adequate and its
linearization yields 
\begin{equation}
 \dot{\varepsilon} = -2\lambda \cos \theta \varepsilon
\mbox{.}
\end{equation}
Assuming that the loss of polarization remains small enough, we can use the
semiclassical ($r=1$) solution~\cite{PY07} of \eqref{eq:semiclassical2}   
\begin{equation}
 \tan\left[\theta(t)/2\right]=e^{-\lambda t}\tan\left[\theta(0)/2\right]
\label{eq:sc-evo-angle}
\end{equation}
that shows that the gyroscope aligns itself
with the axis of polarization of the incoming particle at a rate given by the
semiclassical ratio $\lambda$. Injecting this in the linearized equation of
motion, we can integrate $\cos\theta$ by expressing it in terms of
$\tan\left[\theta(t)/2\right]$ and using \eqref{eq:sc-evo-angle} to turn it into
a rational expression of $e^{-\lambda t}$. The result reads  
\begin{equation}
 \varepsilon (t)=\varepsilon (\tau) \frac{\left( \cosh \lambda \tau +\cos
\theta_0 \sinh  \lambda \tau   \right)^2}{\left( \cosh \lambda t +\cos \theta_0
\sinh  \lambda t   \right)^2}  \mbox{.}
\label{eq:analytical-expression}
\end{equation}

We compare these analytical expressions with numerical simulations. Numerical
results are summarized in Fig. \ref{fig:evo-analytical}; four regimes of
behavior appear, delineated by times $t_1$, $t_2$ and $t_3$. For $t<t_1$,
corrections of higher order than $1/\ell^2$ play an important role and Eq.
\eqref{eq:semiclassical-with-corrections} is not exact, as can be seen in the top
figure. Thus, the initial drop is not captured by semiclassical
equations. However, Eq.
\eqref{eq:semiclassical-with-corrections} accounts
well for the polarization decrease for $t_1<t<t_2$. However, while a formal
solution can be written down by linearizing the equation and using the method of
variation of parameters, we failed to attain a compact analytical formula. We
thus performed a numerical simulation of Eq.
\eqref{eq:semiclassical-with-corrections} which agrees well with the
brute force computation of $\dot{r}$ from the quantum state evolving under the
quantum channel. For the crossover period $t_2<t<t_3$, neither
\eqref{eq:semiclassical-with-corrections} nor \eqref{eq:semiclassical1} accounts
for the evolution of $r$. A better assumption than
\eqref{eq:eigenstate-assumption} is probably needed to properly describe the
transition between \eqref{eq:semiclassical-with-corrections} and
\eqref{eq:semiclassical1}. The return to full polarization for $t>t_3$ is well
 described by the analytical expression
\eqref{eq:analytical-expression} with initial condition $\tau=t_3$.

\section{Decoherence of a superposition of coherent states}

In this section, we will describe the evolution of states that behave
semiclassically, namely coherent states, before investigating the evolution of 
a superposition of such states. It will exhibit dephasing on a much shorter
timescale than relaxation. 

\subsection{Coherent states}

Coherent states have maximal angular momentum along a certain axis. If this axis
is in the $xz$ plane with polar angle $\theta$, the corresponding quantum state
reads, in the $\ket{\ell,m}$ basis,
\begin{equation}
\ket{\theta}=e^{-iL_{y}\theta}|\ell,\ell\rangle\mbox{.}\label{eq:coherent_state}
\end{equation}
The semiclassical equations of motion \eqref{eq:semiclassical1} indicate that
coherent states will
keep maximal polarization ($\forall t\; r(t)\sim1$) and align themselves
with the $z$ axis, rotating around the $y$ axis according to
\eqref{eq:sc-evo-angle}.
Once again, the ratio $\lambda$ plays a crucial role as it sets the
timescale of the alignment of the gyroscope with the $z$ axis, which
is exponentially fast. Equation \eqref{eq:sc-evo-angle} describes the relaxation
or thermalization
of the gyroscope, a classical phenomenon governed by the semiclassical equation
of motions. 
In the next paragraph, we will investigate a purely quantum phenomenon,
dephasing, which typically acts 
on a much shorter timescale.

\subsection{Evolution of a coherent superposition of coherent states}

\begin{figure}
\begin{centering}
\includegraphics[width=0.9\columnwidth]{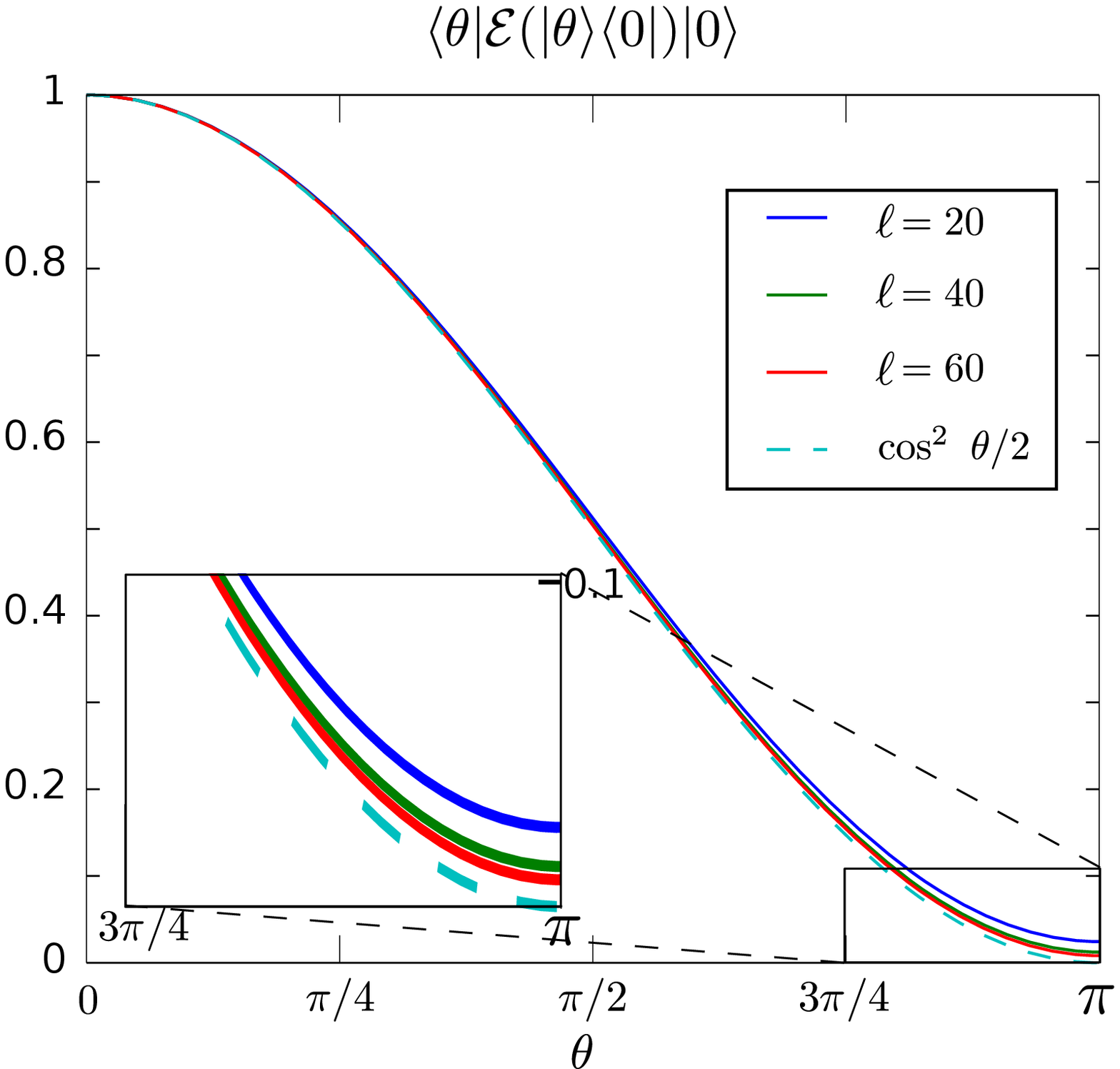}
\end{centering}
\caption{Evolution of the off-diagonal term $\ket{\theta}\bra 0$ corresponding
to the superposition $\alpha\ket{\theta}+\beta\ket 0$ where $\ket 0=\ket{\ell,\ell}$
points in the direction $z$ and $\ket{\theta}$ is given by equation
\eqref{eq:coherent_state}}
\label{fig:evo-superposition}
\end{figure}

Consider a gyroscope initially prepared in a superposition of coherent
states $|\psi\rangle=a|\theta\rangle+b|\phi\rangle$. The evolution of the 
coherence terms can be computed by using the Kraus expression of the quantum
channel
given in \cite{PY07} and expressing operators acting on $\ket{\theta}$ (resp.
$\ket{\phi}$)
in the basis $\{L^\theta_x,\, L^\theta_z\}$ (resp. $\{L^\phi_x,\, L^\phi_z\}$).
The following equation describes how the evolution reduces the coherence terms:\begin{multline}
\langle\theta|\mathcal{E}\left(|\theta\rangle\langle\phi|\right)|\phi\rangle =  \cos^{2}\frac{\theta-\phi}{2}-\frac{\cos\theta-\phi}{d} \\
 +\frac{2\ell\ave{S_{z}}}{d^{2}}\left(\cos\theta+\cos\phi\right)-\frac{2\ave{S_{z}}-\cos^{2}\frac{\theta-\phi}{2}}{d^{2}}\mbox{.} \label{eq:off-diago-evo} \end{multline}
In the semiclassical limit, i.e., $d\gg1$ the first term dominates, as confirmed
by numerical simulations shown in Fig. \ref{fig:evo-superposition} for $\phi=0$.

Thus, the coherence terms vanish more and more rapidly as the initial value
of $|\theta-\phi|$ increases. This is a feature similar to the result
obtained for Quantum brownian motion~\cite{CL83,HPZ92} where the coherence time
of a superposition of two localized wavepackets is inversely
proportional to the square of the distance~\cite{DP97}. The difference of
angles characterizes the \emph{classical distance}
between the two quasi-classical states.

The superposition $|\psi\rangle$ thus evolves according to \begin{multline}
\mathcal{E}\left(|\psi\rangle\langle\psi|\right) = |a|^{2}|\theta+d\theta\rangle\langle\theta+d\theta|+|b|^{2}|\phi+d\phi\rangle\langle\phi+d\phi|\\
 +ab^{*}\left(\cos^{2}\frac{\theta-\phi}{2}|\theta\rangle\langle\phi|+\chi\right)+\mbox{h.c.}
\label{eq:evo-superposition} \end{multline}

\begin{figure}
\begin{centering}
\includegraphics[width=0.9\columnwidth]{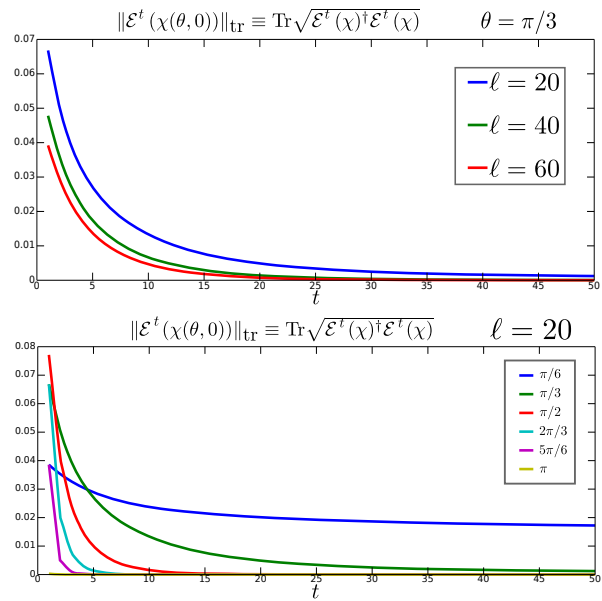}
\par\end{centering}
\caption{Numerical evolution of the trace norm of the operator $\chi$ with different
values of $\ell$ for $\theta=\pi/3$ (top) and with different values
of $\theta$ for $\ell=20$ (bottom). The norm vanishes, indicating
that $\chi$ does not play a role in the evolution of the
superposition of coherent states.}
\label{fig:evo-residual}
\end{figure}

While eq. \eqref{eq:off-diago-evo} gives the expression of a matrix element, eq. \eqref{eq:evo-superposition}
is stronger since it represents a density matrix. However, we had to introduce an operator $\chi$, 
required by the trace-preserving nature of the channel $\mathcal{E}$. This operator accounts for terms 
that do not play a role in subsequent application of the channel evolution in the 
semiclassical limit $\ell \gg 1$ (see Fig. \ref{fig:evo-residual}) as pointed out by numerical evidence shown on Fig. \ref{fig:evo-residual}. 

%

Thus, $|\psi\rangle\langle\psi|$ turns into a mixture $\sim|a|^{2}|\theta(t)\rangle\langle\theta(t)|+|b|^{2}|\phi(t)\rangle\langle\phi(t)|$
after interacting with a few incoming particles. This decoherence process is due to the entanglement
with the incoming particles that allows the environment (made of all the incoming particles) to partially distinguish the two coherent components ($\ket{\theta}$ and $\ket{\phi}$) of the state. In the semiclassical limit $d\gg1$, decoherence acts on a timescale that essentially depends on the ``classical'' distance $|\theta-\phi|$ and \emph{not on the semiclassical ratio $\lambda$} that governs relaxation. Indeed, unless the two coherent components are very close ($\theta \sim \phi$), decoherence acts exponentially fast,
in the sense that $\langle\theta|\mathcal{E}\left(|\theta\rangle\langle\phi|\right)|\phi\rangle$ reduces 
to a value $\epsilon$ after interacting with a number of incoming particles that scales as
$-\log\left(\cos^{2}\frac{\theta-\phi}{2}\right)$. Thus, in general, decoherence occurs on a timescale much shorter than relaxation. Once the two components $\ket{\theta}$ and $\ket{\phi}$ have decohered, they each relax and align with the $z$ axis independantly.
These features are a revealing example of characteristics expected from the general theory of decoherence.

\section{Loss of purity}

The predictability sieve, introduced in~\cite{ZHP93}, identifies semiclassical
states as those which minimize entropy production
or purity loss. In the same way that minimal uncertainty coherent
states are semiclassical for quantum Brownian motion~\cite{ZHP93}, in our
model, the $SU(2)$ coherent states clearly behave semiclassically and should
minimize purity loss.

\begin{figure}
\begin{centering}\medskip
\includegraphics[width=0.9\columnwidth]{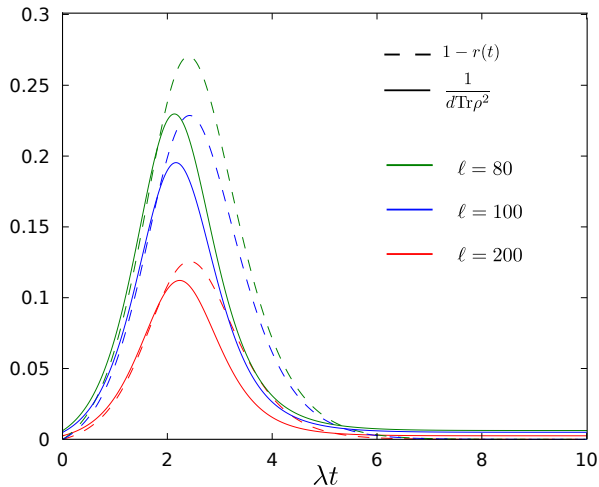}
\par\end{centering}
\caption{Evolution of the  the purity $\mbox{Tr} \rho^2(t)$ for increasing values of $\ell$. Dashed lines represent the loss of polarization $\epsilon=1-r$ and the solid lines represent the inverse of $d\mbox{Tr}\rho^2$. Before the loss of polarization reaches a maximum, solid and dashed lines agree for all values of $\ell$.}
\label{fig:purity}
\end{figure}

Interestingly, they
still undergo a significant purity loss initially. After a few interactions,
the gyroscope will lose a small amount of polarization, i.e.
$r(t)=1-\varepsilon$
where $\varepsilon$ is small for large $l$. The resulting state
$\rho$ will then be roughly a statistical mixture of $\varepsilon d$
states among the $d$ available states. Supposing that each coefficient
has the same order of magnitude, the purity $\mbox{Tr}\rho^{2}$ will
scale as $\varepsilon d(\frac{1}{\varepsilon d})^{2}=\frac{1}{\varepsilon d}$,
explaining the loss of purity. This prediction agrees with numerical
results (Fig. \ref{fig:purity}).

%

\section{Induced POVM}

Interaction between the quantum gyroscope $\mathcal{R}$ and the
spin-$\frac{1}{2}$
particle, which can be represented by a unitary operator $U=e^{-i\mathcal{H}_{\mathcal{RS}}\tau}$,
creates an entangled state. Therefore, a measurement represented by
a POVM $\{\Pi_{\pm}\}$ on the particle yields information about the
state of the gyroscope according to the POVM $\{\Lambda_{\pm}\}$
\begin{equation}
\Lambda_{\pm}=\mbox{Tr}_{\mathcal{S}}\left[(\id_{\mathcal{R}}\otimes\sqrt{\xi}
)U^{\dagger}(\id_{\mathcal{R}}\otimes\Pi_{\pm})U(\id_{\mathcal{R}}\otimes\sqrt{
\xi})\right]
\end{equation}
Our calculations show that a projective measurement of the spin-$\frac{1}{2}$
particle along an axis $\mathbf{u}$ yields 
\begin{equation}
\Lambda_{\pm}=\frac{1}{2}\mbox{Tr}_{\mathcal{S}}\left[\Pi_{\pm}\xi\right]\id_{
\mathcal{R}}\pm\left(\langle\vec{S}\rangle\times\vec{u}\right)\cdot\frac{\vec{L}
}{\ell+\frac{1}{2}}
\end{equation}
corresponding to a measurement of $\mathbf{L}$ along $\langle\mathbf{S}\rangle\times\mathbf{u}$. 

A broader problem is to evaluate the POVM induced by $n$ spin-$\frac{1}{2}$
particles interacting one after the other with the gyroscope during
a time $\tau$ and then measured collectively. It is not clear how
much information about the quantum state $\rho$ of the gyroscope
can be gained this way. It is possible that only information about
a few global properties of $\rho$ can be extracted for large but
finite $n$ thus making these properties good candidates for characteristics
that emerge in the semiclassical limit and which correspond to macroscopic
attributes of the gyroscope.

\subsection*{Acknowledgements}
This work is partially funded by FQRNT and NSERC. We thank David Poulin for
many stimulating discussions about the model and its link to decoherence, and Mychel Pineault for useful discussions.

\bibliography{ref.bib}

\begin{thebibliography}{17}%
\makeatletter
\providecommand \@ifxundefined [1]{%
 \@ifx{#1\undefined}
}%
\providecommand \@ifnum [1]{%
 \ifnum #1\expandafter \@firstoftwo
 \else \expandafter \@secondoftwo
 \fi
}%
\providecommand \@ifx [1]{%
 \ifx #1\expandafter \@firstoftwo
 \else \expandafter \@secondoftwo
 \fi
}%
\providecommand \natexlab [1]{#1}%
\providecommand \enquote  [1]{``#1''}%
\providecommand \bibnamefont  [1]{#1}%
\providecommand \bibfnamefont [1]{#1}%
\providecommand \citenamefont [1]{#1}%
\providecommand \href@noop [0]{\@secondoftwo}%
\providecommand \href [0]{\begingroup \@sanitize@url \@href}%
\providecommand \@href[1]{\@@startlink{#1}\@@href}%
\providecommand \@@href[1]{\endgroup#1\@@endlink}%
\providecommand \@sanitize@url [0]{\catcode `\\12\catcode `\$12\catcode
  `\&12\catcode `\#12\catcode `\^12\catcode `\_12\catcode `\%12\relax}%
\providecommand \@@startlink[1]{}%
\providecommand \@@endlink[0]{}%
\providecommand \url  [0]{\begingroup\@sanitize@url \@url }%
\providecommand \@url [1]{\endgroup\@href {#1}{\urlprefix }}%
\providecommand \urlprefix  [0]{URL }%
\providecommand \Eprint [0]{\href }%
\@ifxundefined \urlstyle {%
  \providecommand \doi  [0]{\begingroup \@sanitize@url \@doi}%
  \providecommand \@doi [1]{\endgroup \@@startlink {\doibase
  #1}doi:\discretionary {}{}{}#1\@@endlink }%
}{%
  \providecommand \doi  [0]{doi:\discretionary{}{}{}\begingroup
  \urlstyle{rm}\Url }%
}%
\providecommand \doibase [0]{http://dx.doi.org/}%
\providecommand \Doi [0]{\begingroup \@sanitize@url \@Doi }%
\providecommand \@Doi  [1]{\endgroup\@@startlink{\doibase#1}\@@Doi}%
\providecommand \@@Doi [1]{#1\@@endlink}%
\providecommand \selectlanguage [0]{\@gobble}%
\providecommand \bibinfo  [0]{\@secondoftwo}%
\providecommand \bibfield  [0]{\@secondoftwo}%
\providecommand \translation [1]{[#1]}%
\providecommand \BibitemOpen [0]{}%
\providecommand \bibitemStop [0]{}%
\providecommand \bibitemNoStop [0]{.\EOS\space}%
\providecommand \EOS [0]{\spacefactor3000\relax}%
\providecommand \BibitemShut  [1]{\csname bibitem#1\endcsname}%
\bibitem [{\citenamefont {Bartlett}\ \emph {et~al.}(2007)\citenamefont
  {Bartlett}, \citenamefont {Rudolph},\ and\ \citenamefont {Spekkens}}]{BRS07}%
  \BibitemOpen
  \bibfield  {author} {\bibinfo {author} {\bibfnamefont {S.~D.}\ \bibnamefont
  {Bartlett}}, \bibinfo {author} {\bibfnamefont {T.}~\bibnamefont {Rudolph}}, \
  and\ \bibinfo {author} {\bibfnamefont {R.~W.}\ \bibnamefont {Spekkens}},\
  }\Doi {10.1103/RevModPhys.79.555} {\bibfield  {journal} {\bibinfo  {journal}
  {Rev. Mod. Phys.},\ }\textbf {\bibinfo {volume} {79}},\ \bibinfo {pages}
  {555} (\bibinfo {year} {2007})}\BibitemShut {NoStop}%
\bibitem [{\citenamefont {Bartlett}\ \emph {et~al.}(2006)\citenamefont
  {Bartlett}, \citenamefont {Rudolph}, \citenamefont {Spekkens},\ and\
  \citenamefont {Turner}}]{BRS+06}%
  \BibitemOpen
  \bibfield  {author} {\bibinfo {author} {\bibfnamefont {S.~D.}\ \bibnamefont
  {Bartlett}}, \bibinfo {author} {\bibfnamefont {T.}~\bibnamefont {Rudolph}},
  \bibinfo {author} {\bibfnamefont {R.~W.}\ \bibnamefont {Spekkens}}, \ and\
  \bibinfo {author} {\bibfnamefont {P.~S.}\ \bibnamefont {Turner}},\
  }\href@noop {} {\bibfield  {journal} {\bibinfo  {journal} {New Journal of
  Physics},\ }\textbf {\bibinfo {volume} {8}},\ \bibinfo {pages} {58} (\bibinfo
  {year} {2006})}\BibitemShut {NoStop}%
\bibitem [{\citenamefont {Poulin}\ and\ \citenamefont {Yard}(2007)}]{PY07}%
  \BibitemOpen
  \bibfield  {author} {\bibinfo {author} {\bibfnamefont {D.}~\bibnamefont
  {Poulin}}\ and\ \bibinfo {author} {\bibfnamefont {J.}~\bibnamefont {Yard}},\
  }\href@noop {} {\bibfield  {journal} {\bibinfo  {journal} {New Journal of
  Physics},\ }\textbf {\bibinfo {volume} {9}},\ \bibinfo {pages} {156}
  (\bibinfo {year} {2007})}\BibitemShut {NoStop}%
\bibitem [{\citenamefont {Ahmadi}\ \emph {et~al.}(2010)\citenamefont {Ahmadi},
  \citenamefont {Jennings},\ and\ \citenamefont {Rudolph}}]{AJR10}%
  \BibitemOpen
  \bibfield  {author} {\bibinfo {author} {\bibfnamefont {M.}~\bibnamefont
  {Ahmadi}}, \bibinfo {author} {\bibfnamefont {D.}~\bibnamefont {Jennings}}, \
  and\ \bibinfo {author} {\bibfnamefont {T.}~\bibnamefont {Rudolph}},\ }\Doi
  {10.1103/PhysRevA.82.032320} {\bibfield  {journal} {\bibinfo  {journal}
  {Phys. Rev. A},\ }\textbf {\bibinfo {volume} {82}},\ \bibinfo {pages}
  {032320} (\bibinfo {year} {2010})}\BibitemShut {NoStop}%
\bibitem [{\citenamefont {Landon-Cardinal}\ and\ \citenamefont
  {MacKenzie}()}]{LM}%
  \BibitemOpen
  \bibfield  {author} {\bibinfo {author} {\bibfnamefont {O.}~\bibnamefont
  {Landon-Cardinal}}\ and\ \bibinfo {author} {\bibfnamefont {R.}~\bibnamefont
  {MacKenzie}},\ }\href@noop {} {\enquote {\bibinfo {title} {Poster
  presentation at the twelfth workshop on quantum information processing (qip),
  january 2009.}}\ }\BibitemShut {NoStop}%
\bibitem [{\citenamefont {Imamo\ifmmode~\bar{g}\else \={g}\fi{}lu}\ \emph
  {et~al.}(2003)\citenamefont {Imamo\ifmmode~\bar{g}\else \={g}\fi{}lu},
  \citenamefont {Knill}, \citenamefont {Tian},\ and\ \citenamefont
  {Zoller}}]{IKT+03}%
  \BibitemOpen
  \bibfield  {author} {\bibinfo {author} {\bibfnamefont {A.}~\bibnamefont
  {Imamo\ifmmode~\bar{g}\else \={g}\fi{}lu}}, \bibinfo {author} {\bibfnamefont
  {E.}~\bibnamefont {Knill}}, \bibinfo {author} {\bibfnamefont
  {L.}~\bibnamefont {Tian}}, \ and\ \bibinfo {author} {\bibfnamefont
  {P.}~\bibnamefont {Zoller}},\ }\Doi {10.1103/PhysRevLett.91.017402}
  {\bibfield  {journal} {\bibinfo  {journal} {Phys. Rev. Lett.},\ }\textbf
  {\bibinfo {volume} {91}},\ \bibinfo {pages} {017402} (\bibinfo {year}
  {2003})}\BibitemShut {NoStop}%
\bibitem [{\citenamefont {Coish}\ and\ \citenamefont {Loss}(2004)}]{CL04}%
  \BibitemOpen
  \bibfield  {author} {\bibinfo {author} {\bibfnamefont {W.}~\bibnamefont
  {Coish}}\ and\ \bibinfo {author} {\bibfnamefont {D.}~\bibnamefont {Loss}},\
  }\Doi {10.1103/PhysRevB.70.195340} {\bibfield  {journal} {\bibinfo  {journal}
  {Phys. Rev. B},\ }\textbf {\bibinfo {volume} {70}},\ \bibinfo {pages} {1}
  (\bibinfo {year} {2004})},\ ISSN \bibinfo {issn} {1098-0121}\BibitemShut
  {NoStop}%
\bibitem [{\citenamefont {Cucchietti}\ \emph {et~al.}(2005)\citenamefont
  {Cucchietti}, \citenamefont {Paz},\ and\ \citenamefont {Zurek}}]{CPZ05}%
  \BibitemOpen
  \bibfield  {author} {\bibinfo {author} {\bibfnamefont {F.}~\bibnamefont
  {Cucchietti}}, \bibinfo {author} {\bibfnamefont {J.}~\bibnamefont {Paz}}, \
  and\ \bibinfo {author} {\bibfnamefont {W.}~\bibnamefont {Zurek}},\
  }\href@noop {} {\bibfield  {journal} {\bibinfo  {journal} {Phys. Rev. A},\
  }\textbf {\bibinfo {volume} {72}} (\bibinfo {year} {2005})},\ ISSN \bibinfo
  {issn} {1050-2947}\BibitemShut {NoStop}%
\bibitem [{\citenamefont {Landon-Cardinal}\ and\ \citenamefont
  {MacKenzie}(2009)}]{LM09}%
  \BibitemOpen
  \bibfield  {author} {\bibinfo {author} {\bibfnamefont {O.}~\bibnamefont
  {Landon-Cardinal}}\ and\ \bibinfo {author} {\bibfnamefont {R.}~\bibnamefont
  {MacKenzie}},\ }\Doi {10.1103/PhysRevA.80.062319} {\bibfield  {journal}
  {\bibinfo  {journal} {Phys. Rev. A},\ }\textbf {\bibinfo {volume} {80}},\
  \bibinfo {pages} {062319} (\bibinfo {year} {2009})}\BibitemShut {NoStop}%
\bibitem [{\citenamefont {Petta}\ \emph {et~al.}(2008)\citenamefont {Petta},
  \citenamefont {Taylor}, \citenamefont {Johnson}, \citenamefont {Yacoby},
  \citenamefont {Lukin}, \citenamefont {Marcus}, \citenamefont {Hanson},\ and\
  \citenamefont {Gossard}}]{PTJ+08}%
  \BibitemOpen
  \bibfield  {author} {\bibinfo {author} {\bibfnamefont {J.~R.}\ \bibnamefont
  {Petta}}, \bibinfo {author} {\bibfnamefont {J.~M.}\ \bibnamefont {Taylor}},
  \bibinfo {author} {\bibfnamefont {A.~C.}\ \bibnamefont {Johnson}}, \bibinfo
  {author} {\bibfnamefont {A.}~\bibnamefont {Yacoby}}, \bibinfo {author}
  {\bibfnamefont {M.~D.}\ \bibnamefont {Lukin}}, \bibinfo {author}
  {\bibfnamefont {C.~M.}\ \bibnamefont {Marcus}}, \bibinfo {author}
  {\bibfnamefont {M.~P.}\ \bibnamefont {Hanson}}, \ and\ \bibinfo {author}
  {\bibfnamefont {A.~C.}\ \bibnamefont {Gossard}},\ }\Doi
  {10.1103/PhysRevLett.100.067601} {\bibfield  {journal} {\bibinfo  {journal}
  {Phys. Rev. Lett.},\ }\textbf {\bibinfo {volume} {100}},\ \bibinfo {pages}
  {067601} (\bibinfo {year} {2008})}\BibitemShut {NoStop}%
\bibitem [{\citenamefont {Reilly}\ \emph {et~al.}(2008)\citenamefont {Reilly},
  \citenamefont {Taylor}, \citenamefont {Petta}, \citenamefont {Marcus},
  \citenamefont {Hanson},\ and\ \citenamefont {Gossard}}]{RTP+08}%
  \BibitemOpen
  \bibfield  {author} {\bibinfo {author} {\bibfnamefont {D.~J.}\ \bibnamefont
  {Reilly}}, \bibinfo {author} {\bibfnamefont {J.~M.}\ \bibnamefont {Taylor}},
  \bibinfo {author} {\bibfnamefont {J.~R.}\ \bibnamefont {Petta}}, \bibinfo
  {author} {\bibfnamefont {C.~M.}\ \bibnamefont {Marcus}}, \bibinfo {author}
  {\bibfnamefont {M.~P.}\ \bibnamefont {Hanson}}, \ and\ \bibinfo {author}
  {\bibfnamefont {a.~C.}\ \bibnamefont {Gossard}},\ }\Doi
  {10.1126/science.1159221} {\bibfield  {journal} {\bibinfo  {journal}
  {Science},\ }\textbf {\bibinfo {volume} {321}},\ \bibinfo {pages} {817}
  (\bibinfo {year} {2008})},\ ISSN \bibinfo {issn} {1095-9203}\BibitemShut
  {NoStop}%
\bibitem [{\citenamefont {Vink}\ \emph {et~al.}(2009)\citenamefont {Vink},
  \citenamefont {Nowack}, \citenamefont {Koppens}, \citenamefont {Danon},
  \citenamefont {Nazarov},\ and\ \citenamefont {Vandersypen}}]{VNK+09}%
  \BibitemOpen
  \bibfield  {author} {\bibinfo {author} {\bibfnamefont {I.~T.}\ \bibnamefont
  {Vink}}, \bibinfo {author} {\bibfnamefont {K.~C.}\ \bibnamefont {Nowack}},
  \bibinfo {author} {\bibfnamefont {F.~H.~L.}\ \bibnamefont {Koppens}},
  \bibinfo {author} {\bibfnamefont {J.}~\bibnamefont {Danon}}, \bibinfo
  {author} {\bibfnamefont {Y.~V.}\ \bibnamefont {Nazarov}}, \ and\ \bibinfo
  {author} {\bibfnamefont {L.~M.~K.}\ \bibnamefont {Vandersypen}},\ }\Doi
  {10.1038/nphys1366} {\bibfield  {journal} {\bibinfo  {journal} {Nat. Phys.},\
  }\textbf {\bibinfo {volume} {5}},\ \bibinfo {pages} {764} (\bibinfo {year}
  {2009})},\ ISSN \bibinfo {issn} {1745-2473}\BibitemShut {NoStop}%
\bibitem [{\citenamefont {Xu}\ \emph {et~al.}(2009)\citenamefont {Xu},
  \citenamefont {Yao}, \citenamefont {Sun}, \citenamefont {Steel},
  \citenamefont {Bracker}, \citenamefont {Gammon},\ and\ \citenamefont
  {Sham}}]{XYS+09}%
  \BibitemOpen
  \bibfield  {author} {\bibinfo {author} {\bibfnamefont {X.}~\bibnamefont
  {Xu}}, \bibinfo {author} {\bibfnamefont {W.}~\bibnamefont {Yao}}, \bibinfo
  {author} {\bibfnamefont {B.}~\bibnamefont {Sun}}, \bibinfo {author}
  {\bibfnamefont {D.~G.}\ \bibnamefont {Steel}}, \bibinfo {author}
  {\bibfnamefont {A.~S.}\ \bibnamefont {Bracker}}, \bibinfo {author}
  {\bibfnamefont {D.}~\bibnamefont {Gammon}}, \ and\ \bibinfo {author}
  {\bibfnamefont {L.~J.}\ \bibnamefont {Sham}},\ }\Doi {10.1038/nature08120}
  {\bibfield  {journal} {\bibinfo  {journal} {Nature},\ }\textbf {\bibinfo
  {volume} {459}},\ \bibinfo {pages} {1105} (\bibinfo {year}
  {2009})}\BibitemShut {NoStop}%
\bibitem [{\citenamefont {Caldeira}\ and\ \citenamefont
  {Leggett}(1983)}]{CL83}%
  \BibitemOpen
  \bibfield  {author} {\bibinfo {author} {\bibfnamefont {A.~O.}\ \bibnamefont
  {Caldeira}}\ and\ \bibinfo {author} {\bibfnamefont {A.~J.}\ \bibnamefont
  {Leggett}},\ }\Doi {DOI: 10.1016/0378-4371(83)90013-4} {\bibfield  {journal}
  {\bibinfo  {journal} {Physica A: Statistical and Theoretical Physics},\
  }\textbf {\bibinfo {volume} {121}},\ \bibinfo {pages} {587 } (\bibinfo {year}
  {1983})},\ ISSN \bibinfo {issn} {0378-4371}\BibitemShut {NoStop}%
\bibitem [{\citenamefont {Hu}\ \emph {et~al.}(1992)\citenamefont {Hu},
  \citenamefont {Paz},\ and\ \citenamefont {Zhang}}]{HPZ92}%
  \BibitemOpen
  \bibfield  {author} {\bibinfo {author} {\bibfnamefont {B.~L.}\ \bibnamefont
  {Hu}}, \bibinfo {author} {\bibfnamefont {J.~P.}\ \bibnamefont {Paz}}, \ and\
  \bibinfo {author} {\bibfnamefont {Y.}~\bibnamefont {Zhang}},\ }\Doi
  {10.1103/PhysRevD.45.2843} {\bibfield  {journal} {\bibinfo  {journal} {Phys.
  Rev. D},\ }\textbf {\bibinfo {volume} {45}},\ \bibinfo {pages} {2843}
  (\bibinfo {year} {1992})}\BibitemShut {NoStop}%
\bibitem [{\citenamefont {D\'avila~Romero}\ and\ \citenamefont
  {Pablo~Paz}(1997)}]{DP97}%
  \BibitemOpen
  \bibfield  {author} {\bibinfo {author} {\bibfnamefont {L.}~\bibnamefont
  {D\'avila~Romero}}\ and\ \bibinfo {author} {\bibfnamefont {J.}~\bibnamefont
  {Pablo~Paz}},\ }\Doi {10.1103/PhysRevA.55.4070} {\bibfield  {journal}
  {\bibinfo  {journal} {Phys. Rev. A},\ }\textbf {\bibinfo {volume} {55}},\
  \bibinfo {pages} {4070} (\bibinfo {year} {1997})}\BibitemShut {NoStop}%
\bibitem [{\citenamefont {Zurek}\ \emph {et~al.}(1993)\citenamefont {Zurek},
  \citenamefont {Habib},\ and\ \citenamefont {Paz}}]{ZHP93}%
  \BibitemOpen
  \bibfield  {author} {\bibinfo {author} {\bibfnamefont {W.~H.}\ \bibnamefont
  {Zurek}}, \bibinfo {author} {\bibfnamefont {S.}~\bibnamefont {Habib}}, \ and\
  \bibinfo {author} {\bibfnamefont {J.~P.}\ \bibnamefont {Paz}},\ }\Doi
  {10.1103/PhysRevLett.70.1187} {\bibfield  {journal} {\bibinfo  {journal}
  {Phys. Rev. Lett.},\ }\textbf {\bibinfo {volume} {70}},\ \bibinfo {pages}
  {1187} (\bibinfo {year} {1993})}\BibitemShut {NoStop}%
\end{thebibliography}%
\bibliographystyle{apsrev4-1.bst}

\end{document}